\documentclass[nofootinbib,aps,twocolumn,showpacs,amsmath,amssymb,unsortedaddress]{revtex4-1}
\usepackage{amssymb}
\usepackage{amsmath}
\usepackage{epsfig}
\usepackage{amsfonts}
\usepackage{color}
\usepackage{float}
\usepackage{latexsym}
\usepackage{mathrsfs}
\usepackage{feynmf}
\usepackage{balance}

\twocolumngrid{\newpage}

\begin{document}
\title{ Hardy's argument and successive spin-s measurements }
\author{Ali Ahanj}
\email{ahanj@ipm.ir}
\affiliation{Department of Physics, Khayyam Institute of Higher Education,  Mashhad, Iran}
\affiliation{School of Physics, Institute for Research in Fundamental Science(IPM), P. O. Box 19395-5531, Tehran, Iran.}
\begin{abstract}
We consider a hidden-variable theoretic description of successive
measurements of non commuting spin observables on a input spin-s
state. In this scenario, the hidden-variable theory leads to a Hardy-type argument that
quantum predictions violate it. We show that the maximum probability of
success of Hardy's argument in quantum theory is
$(\frac{1}{2})^{4s}$, which is more than in the spatial case.
\end{abstract}
\maketitle
\section{INTRODUCTION}
The well-known contradiction between quantum theory and local
realism theory  was first pointed out by Einstein, Podolosky and,
Rosen \cite{epr}. In 1964 John Bell \cite{bell64}  derived an
inequality using local hidden-variable theories ( LHVTs\cite{laudisa}; which could be construed as a form of local realism ) and
showed that quantum mechanics and LHVTs predicted statistically different experimental results for certain combination of correlations in two distant systems. According to Bell's theorem it follows that any classical imitation of quantum mechanics (QM) is necessary nonlocal. Recently,  Gr\"{o}blacher
\emph{et al} \cite{nature}, based on Leggett's inequality \cite{leggett},
have shown  that a broad class of  nonlocal hidden-variable theories (HVTs) is still inconsistent with
QM's predictions in both theory and experiment (for more discussion, see \cite{laudisa}).\\In contrast, some investigators were trying to show a direct
contradiction (without using inequality) of QM versus LHVT
\cite{ghz,hewwood,svetlinchi}. All proofs required a
minimum total of six dimensions in Hilbert space rather than the
four required by Bell in his proof. Hardy
\cite{hardy1,hardy2} gave a proof of nonlocality without
using inequality for two qubits and all pure entangled states except
maximally entangled states. Hardy's argument has been generalized
for many qubits and higher dimensional bipartite systems
\cite{kar,ghosh,jose,clifton}. Cabello gave
another argument of Bell's theorem without inequalities for GHZ and
W states \cite{cabello}. Based on Cabello's logic structure, Liang
\emph{et.al} \cite{liang} provided an example of a two-qubit mixed state
which shows nonlocality, still without inequality. In this sense,
Hardy's logical structure is a special case of Cabello's structure.
In contrast, the authors of Ref. \cite{ahanj2} have proved that
the nonlocality argument proposed by Cabello is more general than
Hardy's nonlocality argument. They showed that the maximum
probability of success of the Hardy and Cabello nonlocality (
for the two-qubit system ) in QM is $0.09$ and
$0.1078$ respectively.\\

Entanglement in time is not introduced in QM because
of the different roles that time and space play  in quantum theory. The
meaning of locality in time is that the results of measurement at
time $t_2$ are independent of any measurement performed at some
earlier time $t_1$ or later time $t_3$. In Ref. \cite{ahanj}, the authors
proposed and analyzed a particular scenario to account for the
deviations of QM from ``realism,'' which involves correlations in the
outputs of successive measurements of noncommuting operations in a
spin-s state. The correlations of successive measurements have been
used by Popescu \cite{popescu} in the context of nonlocal quantum
correlations, to analyze a class of Werner states
\cite{werner}, which are entangled but do not break bipartite
Bell-type inequality. Leggett and Garg \cite{garg} have used
consecutive measurements to challenge the applicability of QM to
macroscopic phenomena (see also \cite{peres}). Bell-type
inequalities with successive measurements
 were first considered by Brunkner \emph{et al}.\cite{brukner}. They have derived CHSH- type inequality \cite{chsh}
 for two successive measurements on an arbitrary state of a single
 qubit and have shown that every such state would violate that
 inequality for proper choice of the measurement setting.\\
 In the present paper we study Hardy's nonlocality arguments for
 the correlations between the outputs of $n$ successive measurements for
 all  $s$-spin  measurements. We show that the maximum probability of success of Hardy's argument
 in the successive measurement is much higher than the spatial ones in a certain
 sense.\\ The paper is organized as follows. In Sec. II, we consider
 the basic scenario in detail. Section III explains the logical
 structure of Hardy's argument on  time locality. In Sec. IV, we show that no time-local stochastic HVT (SHVT) can simultaneously satisfy  Hardy's argument. The maximum
 probability of success of Hardy's argument for n-successive
 spin-$s$ measurements is given in Sec. V. Section VI reports
 the conclusion. A proof is given in the Appendix.
 \section{THE BASIC SCENARIO}
 Consider the following sequence of measurements. A quantum particle
with spin $s$, prepared in the initial state $\rho_0$, is sent
through a cascade of Stern-Gerlach (SG) measurements for the spin
components along the directions given by the unit vectors $\hat{
a_1}, \hat{a_2}, \hat{a_3}, \ldots, \hat{a_n}$ ({\it i.e.},
measurement of observables of the form $\vec{S}.\hat{a}$, where
$\vec{S} = (S_x, S_y, S_z)$ is the vector of spin angular momentum
operators $S_x$, $S_y$, $S_z$ and $\hat{a}$ is a unit vector from
$\mathbb{R}^3$). Each measurement has $2s+1$ possible outcomes. For
the $i$-th measurement, we denote these outcomes (eigenvalues)
$\alpha_i \in \{s, s-1, \ldots, -s\}$. \\
Each of the $(2s+1)^n$ possible outcomes which one gets after
performing $n$ consecutive measurements corresponds to a particular
combination of the results of the measurements at previous $n - 1$
steps and the result of the measurement at the $n$th step. The
probability of each of these $(2s + 1)^n$ outcomes is the joint
probability for such combinations. Note that even though the spin
observables $\vec{S}\cdot\hat{a}_1$, $\vec{S}\cdot\hat{a}_2$,
$\ldots$, $\vec{S}\cdot\hat{a}_n$, whose measurements are being
performed at times $t_1$, $t_2$, $\ldots$, $t_n$, respectively (with
$t_1 < t_2 < \ldots < t_n$), do not commute, the aforementioned joint
probabilities for the outcomes are well defined because each of
these spin observables acts on different states \cite{fine82,anderson05,ballentine}. We emphasize that this is the
joint probability for the results of $n$ actual measurements and not
a joint probability distribution for hypothetical simultaneous
values of $n$ noncommuting observables. Moreover, various sub-beams
( i.e., wave functions) emerging from every SG
apparatus [corresponding to $(2s+1)$ outcomes] at every stage of
measurement are separated, without any overlap or recombination
between them. In other words, the eigen wave packet ${\psi}_{s - j,
t_i, \hat{a}_i} (x)$, corresponding to the eigenvalue $s - j$ of
the observable $\vec{S}\cdot\hat{a}_i$, measured at time $t_i$, will
not have any part in  regions where the SG setups, for
measurement of the observables $\vec{S}\cdot\hat{a}_{i + 1, s }$,
$\vec{S}\cdot\hat{a}_{i + 1, s - 1}$, $\ldots$,
$\vec{S}\cdot\hat{a}_{i + 1, s - j + 1}$, $\vec{S}\cdot\hat{a}_{i +
1, s - j - 1}$, $\ldots$, $\vec{S}\cdot\hat{a}_{i + 1, - s}$, are
situated. We further assume that, between two successive
measurements, the spin state does not change with time, that is,
$\vec{S}$ commutes with the interaction Hamiltonian, if any. Also,
throughout the string of measurements, no component ( i.e.,
sub-beam) is blocked. It should be mentioned here that the time of
each of the measurements  is measured by a common clock.\\
Let us consider an SHVT
consists of: (i) a set $\Lambda$ whose elements $\lambda$ are called
hidden variables;
 (ii) a normalized and positive probability distribution $\rho(\lambda)$ defined on $\Lambda$; and
(iii) a set of probability distributions
$p_{\lambda}\left(s.\hat{a}_1=\alpha_1, s.\hat{a}_2=\alpha_2,\ldots, s.\hat{a}_n=\alpha_n\right)$
 for the outcomes of $n$ successive measurements, such that:
\begin{align}
\label{1} p_{QM}&\left(s.\hat{a}_1=\alpha_1,
s.\hat{a}_2=\alpha_2,\ldots, s.\hat{a}_n=\alpha_n\right)
=\nonumber\\&\int_{\Lambda}d\lambda
\rho(\lambda)p_{\lambda}\left(s.\hat{a}_1=\alpha_1,
s.\hat{a}_2=\alpha_2,\ldots,
s.\hat{a}_n=\alpha_n\right).
\end{align}
Here the quantities on
the left-hand side are the probability distributions which QM attaches to the successive outcomes ${\alpha_1,
\alpha_2,\ldots, \alpha_n}$
 of the considered measurements on spin s.\\ By using the generalized equation (\ref{appendix1}) (see the Appendix), we get:
\begin{align}
p_{QM}&\left(s.\hat{a}_1=\alpha_1, \ldots,
s.\hat{a}_n=\alpha_n\right)=\nonumber\\& [ d^{(s)}_{\alpha_0
\alpha_1}(\beta_1-\beta_0)d^{(s)}_{\alpha_1 \alpha_2}
(\beta_2-\beta_1)\ldots d^{(s)}_{\alpha_{n-1}
\alpha_n}(\beta_n-\beta_{n-1})]^2,
\end{align}
where $d^{(s)}_{\alpha m}(\beta)\equiv\left\langle
s,\alpha|exp(\frac{-iS_y(\beta)}{\hbar})|s,m\right\rangle$ .\\
A deterministic hidden-variable model is a particular instance of a
stochastic one where all probabilities $p_\lambda$ can take only the
values 0 or 1. We now analyze the consequences of SHVT for our
scenario. In general, the outputs of $k$th and $l$th experiments may
be correlated so that
\begin{align}
\label{(3.3)} p(s.\hat{a}_l=\alpha_l \& s.\hat{a}_k=\alpha_k )\neq
p(s.\hat{a}_l=\alpha_l)p(s.\hat{a}_k=\alpha_k).
\end{align}
However, in SHVT we suppose that these correlations have a common
cause represented by a stochastic hidden variable $\lambda$ so that
\begin{align}
\label{(3.4)}
 p_{\lambda}(s.\hat{a}_l=\alpha_l \& s.\hat{a}_k=\alpha_k )=
  p_{\lambda}(s.\hat{a}_l=\alpha_l)p_{\lambda}(s.\hat{a}_k=\alpha_k).
\end{align}
 As a consequence of Eq. (\ref{(3.4)}), the probability
$p_\lambda(s.\hat{a}_k=\alpha_k)$ obtained in a
measurement(say, $\vec{S}\cdot\hat{a}_k$ ) performed at time $t_k$ is
independent of any other measurement (say,$\vec{S}\cdot\hat{a}_l$ )
made at some earlier or later time $t_l$. This is called locality in
time \cite{garg,brukner}.\\One should note that for a two-dimensional QM system, one can always assign values
( deterministically or probabilistically ) to the observables with
the help of a HVT. Once the measurement is done, the system will be
prepared in an output state ( namely, an eigenstate of the
observable), and the earlier HVT may or may not work to reproduce
the values of the observables to be measured in that output state (
prepared after the first measurement). In the present paper, we have
considered the possibility of the existence of an HVT for every input qubit
state which can reproduce the measurement outcomes of $n$ successive
measurements.
\section{THE LOGICAL STRUCTURE OF  HARDY'S  ARGUMENT ON  TIME LOCALITY}
Consider four yes/no-type events $A, A^{'}, B$ and $B^{'}$, where $A$ and $A^{'}$ may happen at
 time $t_1$, and $B$ and $B^{'}$ may happen at another time, $t_2$ ($t_2>t_1$).
 The joint probability that, at the first time ($t_1$), $A$ and, at the second time ($t_2$), B are `` yes" is $0$. The joint probability that, at the first time ($t_1$), $A$ is ``no" and, at the second time ($t_2$), $B^{'}$ is ``yes" is $0$. The joint probability that, at the first time ($t_1$), $A^{'}$ is ``yes" and, at the second time ($t_2$), $B$ is ``no", is $0$. The joint probability that both $A^{'}$ and $B^{'}$ are ``yes" is nonzer. We can write this as follows
\begin{align}
p&(A=+1,B=+1)= 0,\nonumber\\
p&(A=-1,B^{'}=+1)=0,\nonumber\\
p&(A^{'}=+1,B=-1)=0,\nonumber\\
p&(A^{'}=+1,\acute{B}=+1)=p \neq 0.
\end{align}
In the next section, we show that these four statements are not compatible with time-local realism.
 The nonzero probability appearing in the argument is the measure of violation of time-local realism.
  It is interesting that two successive s-spin measurements violate time-local realism.
\section{HARDY'S ARGUMENT FOR $n$ SUCCESSIVE MEASUREMENTS FOR ALL SPIN-$s$ MEASUREMENTS}
We deal with the case where the input state is a pure state whose eigenstates
coincide with those of $s.\hat{a}_0$ for some $\hat{a}_0$ whose
eigenvalues we denote  $\alpha_0=j$. Hardy's argument for a
system of $n$ successive spin-$s$ measurements, in it's minimal form
\cite{minimal}, is given by following conditions:
\begin{equation}
\label{q=0} p(s.\hat{a}_1=j, s.\hat{a}_2=j,\ldots,s.\hat{a}_n=j)=0,
\end{equation}
\begin{align}
\label{h1}
p&(s.\hat{a}_1=j-1, s.\hat{a}^{'}_2=j,\ldots,s.\hat{a}^{'}_n=j)=0,\nonumber\\
p&(s.\hat{a}_1=j-2, s.\hat{a}^{'}_2=j,\ldots,s.\hat{a}^{'}_n=j)=0,\nonumber\\
.&\nonumber\\
.&\nonumber\\
p&(s.\hat{a}_1=-j, s.\hat{a}^{'}_2=j,\ldots,s.\hat{a}^{'}_n=j)=0,\nonumber\\
\end{align}
\begin{equation*}
.
\end{equation*}
\begin{equation*}
.
\end{equation*}
\begin{equation*}
.
\end{equation*}
\begin{align}
\label{hl}
p&(s.\hat{a}^{'}_1=j,\ldots s.\hat{a}_l=j-1,\ldots,s.\hat{a}^{'}_n=j)=0, \nonumber\\
p&(s.\hat{a}^{'}_1=j,\ldots, s.\hat{a}_l=j-2,\ldots,s.\hat{a}^{'}_n=j)=0,\nonumber\\
.&\nonumber\\
.&\nonumber\\
p&(s.\hat{a}^{'}_1=j, \ldots s.\hat{a}_l=-j,\ldots, s.\hat{a}^{'}_n=j)=0,\nonumber\\
\end{align}
\begin{equation}
.\nonumber\\
\end{equation}
\begin{equation}
.\nonumber\\
\end{equation}
\begin{equation}
.\nonumber\\
\end{equation}
\begin{align}
\label{hn}
p&(s.\hat{a}^{'}_1=j, s.\hat{a}^{'}_2=j,\ldots,s.\hat{a}_n=j-1)=0,\nonumber\\
p&(s.\hat{a}^{'}_1=j, s.\hat{a}^{'}_2=j,\ldots,s.\hat{a}_n=j-2)=0,\nonumber\\
.&\nonumber\\
.&\nonumber\\
p&(s.\hat{a}^{'}_1=j, s.\hat{a}^{'}_2=j,\ldots,s.\hat{a}_n=-j)=0,\nonumber\\
\end{align}
\begin{align}
\label{hp}
p(s.\hat{a}^{'}_1=j, s.\hat{a}^{'}_2=j,\ldots,s.\hat{a}^{'}_n=j)=p.
\end{align}
First, we prove here that all time-local SHVTs predict $p=0$. Suppose that a time-local SHVT reproducing, in accordance with Eq.(\ref{(3.4)}), the
quantum predictions exist. Accordingly, if we consider, for example, Eq.(\ref{q=0}), we must have
\begin{align}
\label{x} p&(s.\hat{a^{'}}_1=j,\ldots
,s.\hat{a}_l=j-1,\ldots,s.\hat{a^{'}}_n=j)\notag \\
&=\int_{\Lambda}d\lambda\rho(\lambda)p_{\lambda}(s.\hat{a^{'}}_1=j,\ldots
s.\hat{a}_l=j-1,
\ldots,s.\hat{a^{'}}_n=j)\notag\\&=\int_{\Lambda}d\lambda\rho(\lambda)p_{\lambda}(s.\hat{a^{'}}_1=j)\ldots
p_\lambda (s.\hat{a}_l=j-1)\ldots p_{\lambda}(s.\hat{a^{'}}_n=j)\notag\\&=0,
\end{align}
where the second equality is implied by the time-locality condition of Eq.(\ref{(3.4)}). The last equality in Eq.(\ref{x}) can be fulfilled if and only if the product $p_{\lambda}(s.\hat{a^{'}}_1=j)\ldots p_{\lambda}(s.\hat{a^{'}}_n=j)$ vavishes every time within $\Lambda$. An equivalent result holds for Eqs.(\ref{q=0}-\ref{h1}-\ref{hl}-\ref{hn}-\ref{hp}), leading to:
\begin{align}
\label{qhvt=0} p_{\lambda}(s.\hat{a}_1=j)p_{\lambda}(
s.\hat{a}_2=j)\ldots p_{\lambda}(s.\hat{a}_n=j)=0,
\end{align}
\begin{align}
\label{h1hvt}
p_{\lambda}&(s.\hat{a}_1=j-1) p_{\lambda}(s.\hat{a^{'}}_2=j)\ldots p_{\lambda}(s.\hat{a^{'}}_n=j)=0,\nonumber\\
p_{\lambda}&(s.\hat{a}_1=j-2)p_{\lambda}(s.\hat{a^{'}}_2=j)\ldots p_{\lambda}(s.\hat{a^{'}}_n=j)=0,\nonumber\\
.&\nonumber\\
.&\nonumber\\
p_{\lambda}&(s.\hat{a}_1=-j)p_{\lambda}(s.\hat{a^{'}}_2=j)\ldots p_{\lambda}(s.\hat{a^{'}}_n=j)=0,
\end{align}
\begin{equation*}
.
\end{equation*}
\begin{equation*}
.
\end{equation*}
\begin{equation*}
.
\end{equation*}
\begin{align}
\label{hlhvt}
p_{\lambda}&(s.\hat{a^{'}}_1=j)\ldots p_{\lambda}(s.\hat{a}_l=j-1)\ldots p_{\lambda}(s.\hat{a^{'}}_n=j)=0,\nonumber \\
p_{\lambda}&(s.\hat{a^{'}}_1=j)\ldots p_{\lambda}( s.\hat{a}_l=j-2)\ldots p_{\lambda}(s.\hat{a^{'}}_n=j)=0,\nonumber\\
.&\nonumber\\
.&\nonumber\\
p_{\lambda}&(s.\hat{a^{'}}_1=j) \ldots p_{\lambda}(s.\hat{a}_l=-j)\ldots p_{\lambda}(s.\hat{a^{'}}_n=j)=0,
\end{align}
\begin{equation}
.\nonumber\\
\end{equation}
\begin{equation}
.\nonumber\\
\end{equation}
\begin{equation}
.\nonumber\\
\end{equation}
\begin{align}
\label{hnhvt}
p_{\lambda}&(s.\hat{a}^{'}_1=j)p_{\lambda}(s.\hat{a}^{'}_2=j)\ldots p_{\lambda}(s.\hat{a}_n=j-1)=0,\nonumber\\
p_{\lambda}&(s.\hat{a}^{'}_1=j) p_{\lambda}(s.\hat{a}^{'}_2=j)\ldots p_{\lambda}(s.\hat{a}_n=j-2)=0,\nonumber\\
.&\nonumber\\
.&\nonumber\\
p_{\lambda}&(s.\hat{a}^{'}_1=j) p_{\lambda}(s.\hat{a}^{'}_2=j)\ldots p_{\lambda}(s.\hat{a}_n=-j)=0,
\end{align}
\begin{align}
\label{hphvt}
p_{\lambda}&(s.\hat{a}^{'}_1=j) p_{\lambda}(s.\hat{a}^{'}_2=j)\ldots p_{\lambda}(s.\hat{a}^{'}_n=j)=p\neq 0,
\end{align}
where the first $2jn+1$ equations are supposed to hold almost every time within $\Lambda$, while the last equation has to be satisfied in a subset of $\Lambda$ whose measure according to the distribution $\rho(\lambda)$ is nonzero. To prove the more general result that no conceivable time-local SHVT can simultaneously satisfy Eqs.(\ref{qhvt=0})-(\ref{hphvt}), a manipulation of those equations is required.
To this end, let us sum all equations in  each set. We obtain
\begin{align}
\label{AA}
(&1-p_{\lambda}(s.\hat{a}_1=j))\left[p_{\lambda}(s.\hat{a^{'}}_2=j)\ldots p_{\lambda}(s.\hat{a}^{'}_n=j)\right]=0,\nonumber\\
.&\nonumber\\
.&\nonumber\\
(&1-p_{\lambda}(s.\hat{a}_l=j))\left[p_{\lambda}(s.\hat{a}^{'}_1=j)\ldots p_{\lambda}(s.\hat{a}^{'}_n=j)\right]=0,\nonumber\\
.&\nonumber\\
.&\nonumber\\
(&1-p_{\lambda}(s.\hat{a}_n=j))\left[p_{\lambda}(s.\hat{a}^{'}_1=j)\ldots p_{\lambda}(s.\hat{a}^{'}_{n-1}=j)\right]=0.
\end{align}
Now let us partition the set of hidden variables $\Lambda$ and define the following subsets $A_1, A_2, \ldots A_n$, and $B$ as:
\begin{eqnarray}
\label{AB}
A_1&=& \{ \lambda \in \Lambda|p_{\lambda}(s.\hat{a}_{1}=j)=0\},\nonumber\\
.\nonumber\\
.\nonumber\\
A_l&=& \{ \lambda \in \Lambda|p_{\lambda}(s.\hat{a}_{l}=j)=0\},\\
.\nonumber\\
.\nonumber\\
A_n&=& \{ \lambda \in \Lambda|p_{\lambda}(s.\hat{a}_{n}=j)=0\},\nonumber\\
B&=& \Lambda-\{A_1\cup A_2 \cup \ldots \cup A_n\}.
\end{eqnarray}

We have that, for all $\lambda$  belonging to B,
$p_{\lambda}(s.\hat{a}_1=j)p_{\lambda}( s.\hat{a}_2=j)\ldots
p_{\lambda}(s.\hat{a}_n=j) \neq 0$.If set B had a
nonzero measure according to the distribution $\rho$, that is, if
$\int_B d\lambda \rho(\lambda) \neq 0$, there would be violation of
Eq.(\ref{qhvt=0}) and, consequently, of Eq.(\ref{q=0}). Therefore,
 to fuifill Eq.(\ref{qhvt=0}), the set $A_1\cup A_2 \cup
\ldots \cup A_n$ must coincide with $\Lambda$ apart from a set of
zero measure, and we are left only with hidden variables belonging to
either $A_1$ or $A_2$ or $\ldots$ or $A_n$. If $\lambda$ belongs to
$A_l$, then, by definition, $p_{\lambda}(s.\hat{a}_l=j)=0$, so that
Eq.(\ref{AA}) can be satisfied only if
$p_{\lambda}(s.\hat{a^{'}}_1=j)\ldots
p_{\lambda}(s.\hat{a^{'}}_n=j)=0$. Hence, for any $\lambda \in
\{A_1\cup A_2 \cup \ldots \cup A_n \}$, we  obtain a result leading
to a contradiction of  Eq.(\ref{hphvt}), which requires that there
is a set of nonzero $\rho$ measure within $\Lambda$ where both
probabilities do not vanish. To summarize, we have shown that it is
not possible to exhibit any time-local hidden-variable model,
satisfying Hardy's logic for $n$ successive measurements.\\
In the next section, we show that in quantum theory for the $n$ successive spin measurement, sometimes $p>0$.
\section{HARDY'S ARGUMENT RUNS FOR $n$ SUCCESSIVE SPIN MEASUREMENTS BY QUANTUM MECHANICS}
Hardy's nonlocality argument is considered weaker than Bell inequalities in the bipartite case, as every maximally entangled state of two spin-$\frac{1}{2}$ particles violates Bell's inequality maximally but none of them satisfies Hardy-type nonlocality conditions. The scenario in successive spin measurements is quite different, however. We showed in \cite{ahanj} ( also see \cite{brukner}) that all $n$ successive spin-$s$ measurements break Bell-type inequalities, in contrast to the bipartite case, where only the entangled states break it. In this section, we prove that all $n$ successive spin-$s$ measurements satisfy Hardy-type argument conditions.\\ We now consider $n$ successive measurements in directions $s.\hat{a}_{i}$ ($i=1,2,\ldots,n$) on  spin-$s$ particles. For a spin-$s$ system, we have ( see the Appendix):
\begin{equation}\label{pp}
    |\langle \alpha_{k-1}|\alpha_{k} \rangle|=|\langle s.\hat{a}_{k-1}|s.\hat{a}_{k} \rangle|=d_{\alpha_{k-1},\alpha_k}^{(s)}(\beta_k-\beta_{k-1}),
\end{equation}
where $\beta_k$ is the angle between the $\hat{a}_k$ and the $+z$ axes. So, given the input state $|\alpha_0\rangle$, the ( joint) probability that the measurement outcomes will be $|\alpha_1\rangle \in \{+j,\ldots,-j\}$ in the first measurement, $|\alpha_2 \rangle\in \{+j,\ldots,-j\}$ in the second measurement,..., $|\alpha_n\rangle \in \{+j,\ldots,-j\}$ in the $n$-th measurement, is given by
\begin{align}\label{ppp}
    p(\alpha_1,\alpha_2,\ldots,\alpha_n)=&\Pi_{k=1}^{n}|\langle \alpha_{k-1}|\alpha_{k} \rangle|\nonumber\\=&\Pi_{k=1}^{n} d^{2}_{\alpha_{k-1},\alpha_k}(\beta_k-\beta_{k-1}).
\end{align}
We deal with the case where the input state is a pure state whose eigenstates coincide with those of $\vec{S}.\hat{a}_0$ for some $\hat{a}_0$ whose eigenvalues we denote $\alpha_0=j$.  Now, by substituting  Eq.(\ref{ppp}) in the minimal form of  Hardy's argument [ Eqs.(\ref{q=0})-(\ref{hp})],
we have
\begin{align}
\label{q} d^2_{jj}(\beta_1)d^2_{jj}(\beta_2-\beta_1)\ldots
d^2_{jj}(\beta_n-\beta_{n-1})=0,
\end{align}
\begin{align}
\label{c1}
d&^2_{j,j-1}(\beta_1)d^2_{j,j-1}(\beta_2^{'}-\beta_1)\ldots d^2_{jj}(\beta_n^{'}-\beta_{n-1}^{'})=0,\nonumber\\
d&^2_{j,j-2}(\beta_1)d^2_{j,j-2}(\beta_2^{'}-\beta_1)\ldots d^2_{jj}(\beta_n^{'}-\beta_{n-1}^{'})=0,\nonumber\\
.&\nonumber\\
.&\nonumber\\
d&^2_{j,-j}(\beta_1)d^2_{j,-j}(\beta_2^{'}-\beta_1)\ldots d^2_{jj}(\beta_n^{'}-\beta_{n-1}^{'})=0,\nonumber\\
\end{align}
\begin{equation}
.\nonumber
\end{equation}
\begin{equation}
.\nonumber
\end{equation}
\begin{align}
\label{cl}
d&^2_{jj}(\beta^{'}_1)\ldots d^2_{j,j-1}(\beta_l-\beta^{'}_{l-1})d^2_{j,j-1}(\beta^{'}_{l+1}-\beta_{l})\ldots\nonumber \\d&^2_{jj}(\beta^{'}_n-\beta^{'}_{n-1})=0,\nonumber\\
d&^2_{jj}(\beta^{'}_1)\ldots d^2_{j,j-2}(\beta_l-\beta^{'}_{l-1})d^2_{j,j-2}(\beta^{'}_{l+1}-\beta_{l})\ldots\nonumber \\d&^2_{jj}(\beta^{'}_n-\beta^{'}_{n-1})=0,\nonumber\\
.&\nonumber\\
.&\nonumber\\
d&^2_{jj}(\beta^{'}_1)\ldots d^2_{j,-j}(\beta_l-\beta^{'}_{l-1})d^2_{j,-j}(\beta^{'}_{l+1}-\beta_{l})\ldots\nonumber\\ d&^2_{jj}(\beta^{'}_n-\beta^{'}_{n-1})=0,
\end{align}
\begin{equation}
.\nonumber
\end{equation}
\begin{equation}
.\nonumber
\end{equation}
\begin{align}
\label{cn}
d&^2_{jj}(\beta_1^{'})d^2_{jj}(\beta_2^{'}-\beta_1^{'})\ldots d^2_{j,j-1}(\beta_n-\beta_{n-1}^{'})=0,\nonumber\\
d&^2_{jj}(\beta_1^{'})d^2_{jj}(\beta_2^{'}-\beta_1^{'})\ldots d^2_{j,j-2}(\beta_n-\beta_{n-1}^{'})=0,\nonumber\\
.&\nonumber\\
.&\nonumber\\
d&^2_{jj}(\beta_1^{'})d^2_{jj}(\beta_2^{'}-\beta_1^{'})\ldots d^2_{j,-j}(\beta_n-\beta_{n-1}^{'})=0,\nonumber\\
\end{align}
\begin{equation}
\label{p}
d^2_{jj}(\beta_1^{'})d^2_{jj}(\beta_2^{'}-\beta_1^{'})\ldots d^2_{j,j}(\beta_n^{'}-\beta_{n-1}^{'})=p.
\end{equation}
From Eq (\ref{q}), at least one of the factors must be $0$. So
\begin{align*}
d^2_{jj}(\beta_1)=0&\Longrightarrow  \beta_1=\pi\\&\rm{or}\\
d^2_{jj}(\beta_2-\beta_1)&\Longrightarrow |\beta_2-\beta_1|=\pi\\&\rm{or}\\&.\\&.\\&.\\
d^2_{jj}(\beta_n-\beta_{n-1})&\Longrightarrow |\beta_n-\beta_{n-1}|=\pi
\end{align*}
To satisfy all equations (\ref{c1})-(\ref{cn}), we have the following conditions:
\begin{align}
\label{condition}
(\beta_1=0) ~&\rm{or}~ (\beta^{'}_2=\beta_1)\nonumber\\ &\rm{and}\nonumber\\ (\beta_2=\beta^{'}_1)~ &\rm{or}~ (\beta_2=\beta^{'}_3)\nonumber\\&.\nonumber\\&.\nonumber\\&.\nonumber\\(\beta_l=\beta^{'}_{l-1}) &~\rm{or}~ (\beta_l=\beta^{'}_{l+1})\\&.\nonumber\\&.\nonumber\\&.\nonumber\\&\rm{and}\nonumber\\(\beta_n&=\beta^{'}_{n-1}).\nonumber
\end{align}
Now, we can calculate the maximum value $p$ by using these
conditions. For example, if  we select $ \beta_1=\pi$, so we must
have $\beta^{'}_2=\beta_1=\pi$. In this case,
\begin{align}
p=d^2_{jj}(\beta_1^{'})d^2_{jj}(\pi-\beta_1^{'})d^2_{jj}(\beta_3^{'}-\pi)\ldots
d^2_{j,j}(\beta_n^{'}-\beta_{n-1}^{'}).
\end{align}
By substituting $d^{j}_{jj}(\beta)= \cos^{2j}(\beta/2)$, we have
\begin{align}
p&=\cos^{4j}(\frac{\beta_1^{'}}{2})\cos^{4j}(\frac{\pi-\beta_1^{'}}{2})\cos^{4j}(\frac{\beta_3^{'}-\pi}{2})\ldots\nonumber\\ &\cos^{4j}(\frac{\beta_n^{'}-\beta_{n-1}^{'}}{2}).
\end{align}
By selecting $\beta^{'}_n=\beta^{'}_{n-1}=\ldots=\beta^{'}_3=\pi$ and $\beta^{'}_1=\pi$,
\begin{equation}
p\leq(\frac{1}{2})^{4j}.
\end{equation}
We can obtain this result in the general case.
We choose $|\beta_l-\beta_{l-1}|=\pi$, where $2\leq l\leq n$. Without loss of generality, we select $\beta_l=\pi$ and $\beta_{l-1}=0$. From the results obtained with Eq.(\ref{condition}), $\beta^{'}_{l+1}=\beta_l=\pi$ or $\beta^{'}_{l-1}=\beta_l=\pi$. Exactly for (l-1)th in Eq.(\ref{condition}), we have $\beta^{'}_l=\beta_{l-1}=0$ or $\beta^{'}_{l-2}=\beta_{l-1}=0$. So we have four cases: (i)  $\beta^{'}_{l-1}=\pi$ and $\beta^{'}_{l-2}=0$ (ii)$\beta^{'}_{l-1}=\pi$ and $\beta^{'}_{l}=0$ (iii)$\beta^{'}_{l+1}=\pi$ and $\beta^{'}_{l}=0$ and (iv)$\beta^{'}_{l+1}=\pi$ and $\beta^{'}_{l-2}=0$. It is easy see that for the first three cases , the maximum value of $p$ is $0$, but in the forth case, by selecting $\beta^{'}_1=\beta^{'}_2=\ldots=\beta^{'}_{l-1}=0$ and $\beta^{'}_{l+2}=\beta^{'}_{l+3}=\ldots=\beta^{'}_{n}=\pi$, we get
\begin{equation}
p=\cos^{4j}(\frac{\beta^{'}_l}{2})\sin^{4j}(\frac{\beta^{'}_l}{2})
\leq (\frac{1}{2})^{4j}.
\end{equation}
We see that $p>0$ for all spins, and also, the maximum probability of
success of  Hardy's non-time locality is independent of the
number of successive measurements and decreases with $s$.
\section{CONCLUSION}
We considered an HVT description of successive
measurements of noncommuting spin observables on an input spin-$s$
state. Although these spin observables are noncommuting, they act
on different states and so the joint probabilities for the outputs
of successive measurements are well defined. In Ref. \cite{ahanj}, we
account for the maximum violation of these inequalities by quantum
correlations by varying the spin value and the number of successive
measurements. Our approach can be used to obtain a measure of the
deviation of QM from theory obeying realism and
time locality and may lead to a sharper understanding of QM.\\ In
the present paper, we have studied Hardy's argument for the correlations
between the outputs of $n$ successive measurements for all $s$-spin
measurements. We have shown that the maximum probability of success
of Hardy's argument for $n$ successive measurements is $(\frac{1}{2})^{4s}$, which is independent of
the number of successive measurements of spin ($n$) and decreases with increase of
$s$. This can be compared with the
 correlations corresponding to measurement of spin observables in
 a spacelike separated two-particles scenario where only the non-maximally entangled states of any spin-s bipartite system
 respond to  Hardy's nonlocality test. The authors of Ref.
 \cite{minimal} showed that the maximum nonlocalities for a given
 pair of noncommuting spin-$s$ observables per site  turned out
 to be the same ($0.0901099$) for the $s=\frac{1}{2}, 1, and
 \frac{3}{2}$. So the maximum amount of nonlocality that can be
 obtained by  Hardy's nonlocality test on bipartite systems of
 higher spin values also remains the same.
 \section*{ACKNOWLEDGMENTS}
 It is a pleasure to acknowledge Pramod Joag and Guruprasad Kar for
 useful discussions.
\section*{APPENDIX}
Let us consider a situation where an ensemble of systems prepared in
state $\rho=|s.\hat{a}_0=\alpha_0 >< s.\hat{a}_0=\alpha_0|$ at time
$t=0$, is subjected to a measurement of the observable $A(t_1)=
s.\hat{a}_1$ at time $t_1$ followed by a measurement of the
observable $B(t_2)= s.\hat{a}_2$  at time $t_2$ $(t_2>t_1>0)$, where
we have adopted the Heisenberg picture of time evolution. Further,
let us assume that both $A(t_1)$ and $B(t_2)$ have purely discrete
spectra. Let $\{\alpha_1\}=\{-s,-s+1,\ldots,s\}$ and
$\{\alpha_2\}=\{-s,-s+1,\ldots,s\}$ denote the eigenvalues and
$P^{A(t_1)}(\alpha_1)=|s.\hat{a}_1=\alpha_1 ><
s.\hat{a}_1=\alpha_2|$, $P^{B(t_2)}(\alpha_2)=|s.\hat{a}_2=\alpha_2
>< s.\hat{a}_2=\alpha_2|$ the corresponding eigenprojectors of
$A(t_1)$ and $B(t_2)$ respectively. Then the joint probability that
a measurement of $A(t_1)$ yields the outcome $\alpha_1$ and a
measurement of $B(t_2)$ yields the outcome $\alpha_2$ is given by 
\begin{align}
\label{ABC} Pr&^{\rho}_{A(t_1),B(t_2)}(\alpha_1,\alpha_2)\nonumber\\&= Tr
\left[P^{B(t_2)}(\alpha_2)P^{A(t_1)}(\alpha_1)\rho
P^{A(t_1)}(\alpha_1)P^{B(t_2)}(\alpha_2)\right]\nonumber\\&=\left|\left\langle
s.\hat{a}_0 | s.\hat{a}_1\right\rangle\right|^2 \left|\left\langle
s.\hat{a}_1 | s.\hat{a}_2\right\rangle\right|^2.
\end{align}
We know that $|s.a=\alpha >=\sum^{+s}_{m=-s}
d^{(s)}_{\alpha,m}(\beta)|m>$
 where $d^{(s)}_{\alpha m}(\beta)\equiv\left\langle s,\alpha|exp(\frac{-iS_y(\beta)}{\hbar})|s,m\right\rangle$
 and it obtains  Wigner's formula \cite{sakurai} and
  $ \alpha_i \in {-s,\ldots,s}$ and $\beta_i$ is the angle between the $\hat{a}_i$ and the $z$ axes.
In contrast,
\begin{align}
\label{appendix} \left|\left\langle s.\hat{a}_1 |
s.\hat{a}_2\right\rangle\right|&=\sum_{m}
<m|d^{*(s)}_{\alpha_1,m}(\beta_1)\sum_{m^{'}}
d^{(s)}_{\alpha_2,m^{'}}(\beta_2)|m^{'}>\nonumber\\&=\sum_{m}d^{*(s)}_{\alpha_1,m}(\beta_1)d^{(j)}_{\alpha_2,m^{'}}(\beta_2)\nonumber\\
&=d^{(s)}_{\alpha_1\alpha_2}(\beta_2-\beta_1).
\end{align}
So, we obtain
\begin{equation}
\label{appendix1} \rm pr_{QM}(\alpha_1,\alpha_2)=
|d^{(s)}_{\alpha_0\alpha_1}(\beta_1-\beta_0)|^{2}|d^{(s)}_{\alpha_1\alpha_2}(\beta_2-\beta_1)|^{2}.
\end{equation}

\end{document}